# Controllable p–n junctions in three-dimensional Dirac semimetal Cd$_3$As$_2$ nanowires

Janice Ruth Bayogan,[†,‡,¶] Kidong Park,[§,¶] Zhuo Bin Siu,[∥,¶] Sung Jin An,[†] Chiu-Chun Tang,[⊥] Xiao-Xiao Zhang,[#,▽] Man Suk Song,[†,‡] Jeunghee Park,[§] Mansoor B. A. Jalil,[∥] Naoto Nagaosa,[#,@] Kazuhiko Hirakawa,[*,⊥] Christian Schönenberger,[*,△] Jungpil Seo[*,†,‡] and Minkyung Jung[*,‡]

[†] *Department of Emerging Materials Science, DGIST, Daegu 42988, Korea*

[‡] *DGIST Research Institute, DGIST, Daegu 42988, Korea*

[§] *Department of Chemistry, Korea University, Sejong 339-700, Korea*

[∥] *Electrical and Computer Engineering, National University of Singapore, Singapore 117576, Republic of Singapore*

[⊥] *Institute for Nano Quantum Information Electronics, IIS, The University of Tokyo, Tokyo 153-8505, Japan*

[#] *Department of Applied Physics, The University of Tokyo, Tokyo 113-8656, Japan*

[▽] *Department of Physics and Astronomy & Stewart Blusson Quantum Matter Institute, University of British Columbia, Vancouver, BC, V6T 1Z4, Canada*

[@] *RIKEN Center for Emergent Matter Science (CEMS), Saitama 351-0198, Japan*

[△] *Department of Physics, University of Basel, Klingelbergstrasse 82, CH-4056 Basel, Switzerland*

[¶] *J.R.B., K.P. and Z.B.S contributed equally to this work.*

E-mail: minkyung.jung@dgist.ac.kr; jseo@dgist.ac.kr; Christian.Schoenenberger@unibas.ch; hirakawa@iis.u-tokyo.ac.jp

**Abstract**

We demonstrate a controllable p−n junction in a three-dimensional Dirac semimetal (DSM) $Cd_3As_2$ nanowire with two recessed bottom gates. The device exhibits four different conductance regimes with gate voltages, the unipolar (n−n and p−p) regime and the bipolar (n−p and n−p) one, where p−n junctions are formed. The conductance in the p−n junction regime decreases drastically when a magnetic field is applied perpendicular to the nanowire, which is due to the suppression of Klein tunneling. In this regime, the device shows quantum dot behavior. On the other hand, clear conductance plateaus are observed in the n−n regime likely owing to the cyclotron motion of carriers at high magnetic fields. Our experiment shows that the ambipolar tunability of DSM nanowires can enable the realization of quantum devices based on quantum dots and electron optics.

# Keywords

Three-dimensional (3D) Dirac semimetals (DSMs) such as $Cd_3As_2$ and $Na_3Bi$ have attracted considerable attention owing to their exotic electronic properties.[1-4] Unlike topological insulators which have a bulk energy gap, in DSMs the bulk valence and conduction band touch at discrete points in momentum space, known as Dirac nodes. Around the Dirac nodes, the energy-dispersion relation of the electron states is linear along all three momentum directions. The presence of the Dirac nodes is protected by crystal symmetry, establishing DSMs as a 3D analogue of graphene.[1-4] The unique energy band structure ensures that DSMs are not only a source of many intriguing physical phenomena but also a new potential platform for quantum devices.[5-10]

A number of novel electron transport features have been discovered in 3D DSMs, the most striking of which is the so-called chiral anomaly.[11-13] The application of parallel electric and magnetic fields produces an additional axial current, resulting in a negative magnetoresistance (MR). In contrast, application of a magnetic field perpendicular to the electric field leads to a gigantic positive MR.[14-17] Bulk DSMs exhibit ultra-high carrier mobilities with values as high as $10^6$ $cm^2/Vs$.[14] Recent research led to the observation of the quantum Hall effect in DSM $Cd_3As_2$ nano-platelets[18-19] and thin films.[20-24] Moreover, Aharonov-Bohm (AB) oscillations in $Cd_3As_2$, which indicates the ballistic transport dominated by the surface states of DSMs nanowires, have been demonstrated.[25] In DSM nanowires, quantum dot devices have been demonstrated in an unintentionally formed n–p–n cavity with high magnetic fields by suppressing Klein tunneling.[26] As the bands in DSMs are linear and gapless in the bulk, one expects to be able to demonstrate controllable p−n junction devices with two gates as performed in graphene devices,[27] which has not been done yet.

Here, we fabricate a suspended 3D DSM nanowire device with two recessed bottom gates and characterize the p–n junction properties in the presence of a magnetic field. The device exhibits four different conductance regimes as a function of the two gate voltages: a unipolar n−n and p−p regimes and the bipolar n−p and p−n ones, confirming that the device forms a p−n junction. In the bipolar regime, we observe strong conductance suppression at high magnetic fields owing to the suppression of Klein tunneling and confinement effect. On the other hand, we observe conductance quantization in the n−n regime, which is likely due to the quantum edge states that develop in the presence of magnetic fields.

Cd$_3$As$_2$ nanowires are grown by the chemical vapor deposition method. Cd$_3$As$_2$ (99%, Alfa Aesar) powder is placed inside a quartz tube reactor. The Bi catalytic nanoparticles for growth are formed on a silicon substrate by coating 1 mM BiI$_3$ (99.999%, Sigma-Aldrich Corp.) in ethanol solution. The substrate is located 8 cm away from the powder source. Argon gas is continuously supplied at a rate of 200 sccm under ambient pressure during growth. The temperature of the powder sources is set to 450 ºC, while that of the substrate is maintained at $T \approx 350$ ºC for the nanowire growth. Figure 1(a) shows a representative transmission electron microscopy (TEM) image of a Cd$_3$As$_2$ nanowire, which confirms its single crystalline nature. The TEM image shows several amorphous layers on the surface of the nanowires. The growth direction of the nanowires is the [112] direction, which is the axial direction. Details of the growth process and characterization of the nanowires are outlined in Ref. 26.

To realize clean and tunable p–n junction devices in Cd$_3$As$_2$ nanowires, we implemented a suspended device structure with two recessed bottom gates.[28] Figure 1b and c show schematic illustrations of the device and representative scanning electron microscopy (SEM) image with the measurement setup. Suspended Cd$_3$As$_2$ nanowire devices are fabricated on a highly doped p$^{++}$ silicon substrate, which serves as a global backgate, covered by 300 nm-thick thermally grown SiO$_2$. The trenches for the bottom gates are fabricated by electron beam lithography and anisotropic reactive ion etching with CF$_4$, followed by a wet etching step using buffered HF. The depth and width of these trenches are ~100 nm and ~ 300 nm, respectively. Ti/Au (5/25 nm) is deposited in the trenches to form the bottom gates, which are 300 nm wide and spaced at a pitch of 550 nm, as shown in Figure 1(c). Subsequently, Cd$_3$As$_2$ nanowires with the typical diameter of 70 – 110 nm are transferred across the bottom gates (110 nm for Device A). Finally, Ti/Au (5/120 nm) contacts separated by ~ 1.5 μm are defined after etching with Ar plasma to remove the native oxide at the surface of the nanowire. The nanowire is separated from the bottom gates by ~ 70 nm, enabling effective gating compared to the global backgate. Figure 1d shows a sketch of the energy band diagram for each segment of the nanowire (Device A). Independent control of the bottom gates ($V_L$ and $V_R$) enables us to tune the local electronic potential of the nanowire segments above theses gates, shown as a black line in Fig. 1(d). The nanowire leads (indicated in Fig. 1(b)) and nanowire segments underneath the source-drain contacts (denoted by S and D, respectively) are less affected by the bottom gating because of the relatively large distance from the gates. Therefore, these regions effectively remain in the intrinsic doping state, which is weak n–type, as illustrated in Fig. 1(d). The devices are measured using a standard lock-in technique with a modulation frequency of 77 Hz and an excitation voltage of 200 µV at a temperature of ~ 300 mK.

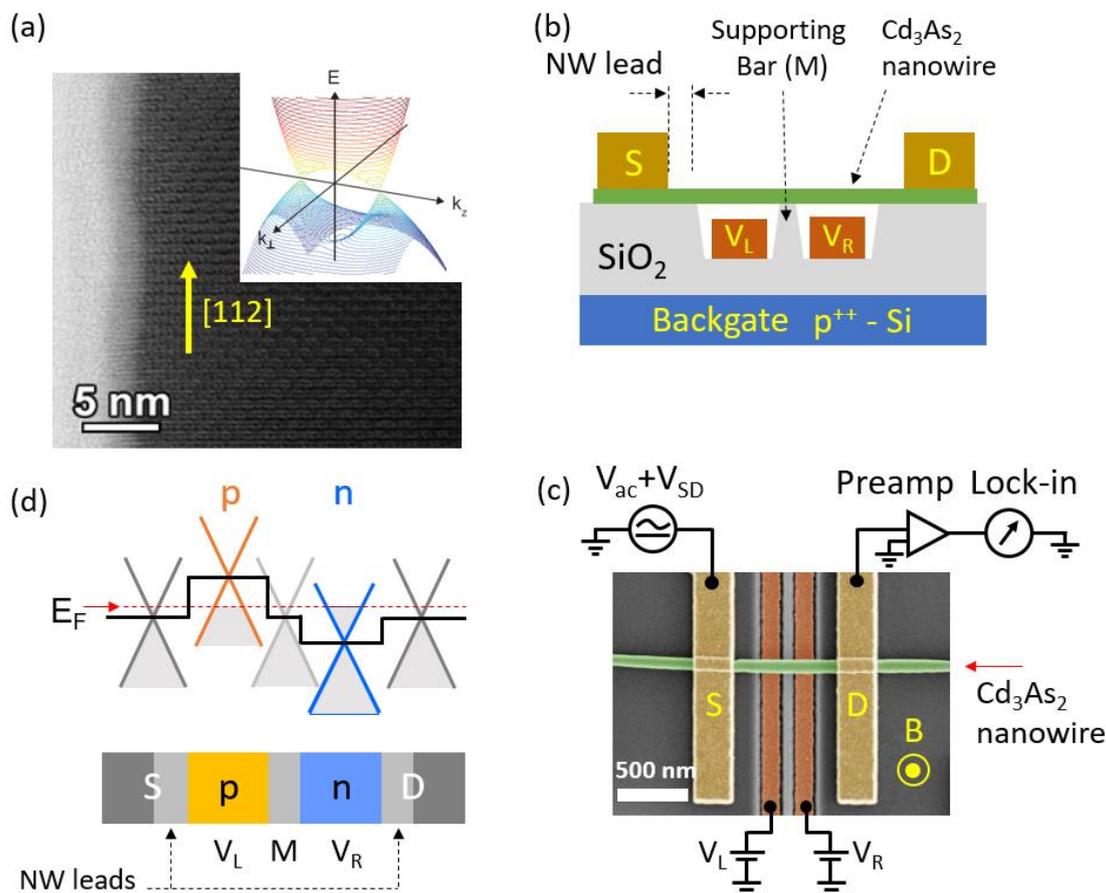

**Figure 1.** (a) Transmission electron microscopy (TEM) image of a Dirac semimetal $Cd_3As_2$ nanowire. The nanowire is grown along the [112] direction. Inset: Schematic of the energy dispersion $E(k)$ near the Dirac nodes of $Cd_3As_2$ based on calculation. (b) Schematic of suspended $Cd_3As_2$ nanowire p–n junction device. The trenches in the $SiO_2$ are fabricated by anisotropic reactive ion etching followed by wet chemical etching. Ti/Au bottom gates (5/25 nm thick) are deposited in the trenches. The nanowire is positioned across the bottom gates and contacted with 10/100 nm of Ti/Au. The $p^{++}$ doped Si substrate acts as a global back gate. (c) Representative scanning electron microscopy (SEM) image and measurement setup of the $Cd_3As_2$ nanowire device. The channel length and diameter of the nanowire for Device A are ~ 1.5 µm and ~ 110 nm, respectively. The magnetic field is applied perpendicular to the nanowire. (d) Schematic of energy band diagram of the p–n junction formed using the two bottom gates. The carrier type (n or p) and carrier density in each segment over the right or left bottom gates

can be controlled independently by adjusting the gate voltage. Because the bottom gates are less effective for the nanowire lead parts around source-drain contacts (denoted as S and D), the potentials in these areas remain at those of the intrinsic doping state (weak n–type for Device A).

Figure 2(a) shows the differential conductance measured as a function of $V_L$ and $V_R$ between the two contacts at $B = 0$ T. Because 3D DSM $Cd_3As_2$ is a 3D analogue of graphene, a conductance plot similar to a graphene p–n junctions is expected. As anticipated, the 2D conductance reveals four different regimes indicated by the labels p–p, n–n, n–p and p–n, which correspond to the different carrier types, over the left and right bottom gates of the device (the first symbol indicates the carrier type of the nanowire segment over the left gate region, whereas the second refers to that over the right gate region). The black dashed lines indicate the charge neutrality point (or Dirac nodes) of the left and right nanowire segments between the unipolar (p–p or n–n) and bipolar regimes (p–n or p–n). The border lines separating the four different regimes are approximately perpendicular to each other, implying that capacitive cross coupling between the two gates is weak. In most suspended devices, the Dirac nodes are observed at approximately zero gate voltage (see Supporting Information, Fig. S1 and S2 for other devices), whereas the Dirac nodes in non-suspended devices are shifted to high negative or positive gate voltages, indicating a n– or p–type behavior, respectively, due to charge traps at the interface between the substrate and nanowires. This fact shows that the suspended structure yields a clean device.

The energy band diagrams for the different doping states are displayed in Fig. 2b. The Dirac nodes appear at slightly negative voltages, indicating that the entire nanowire is weakly n–doped and thus the Fermi energy (red dashed line in Fig. 2b) is slightly above the Dirac nodes. In the n–n regime ((i) in Fig. 2b), all the segments are n–doped, exhibiting high conductance. We can form an n–p–n cavity over the right gate ($V_R$) by applying a negative gate voltage $V_R < 0$, as illustrated in (ii) of Fig. 2b. In the same way, another n–p–n cavity can be formed over the left gate, also by applying a negative voltage $V_L < 0$ (not shown here). When negative voltages are applied to both bottom gates, the device turns into the p–p region (iii). Interestingly, the conductance in the p–p region is much lower than that in the bipolar regions, i.e., p–n and n–p. The reason is twofold. First, the Fermi velocity for holes is approximately three times lower than that for electrons in $Cd_3As_2$.[7] As a result, the hole mobility is much lower than the electron mobility, which results in the asymmetric transfer curve as a function of gate voltage (Supporting Information, Figure S3). Second, double p-n junctions are formed through the nanowire in the p–p region, which further suppress the conductance ((iii) in Fig. 2b).

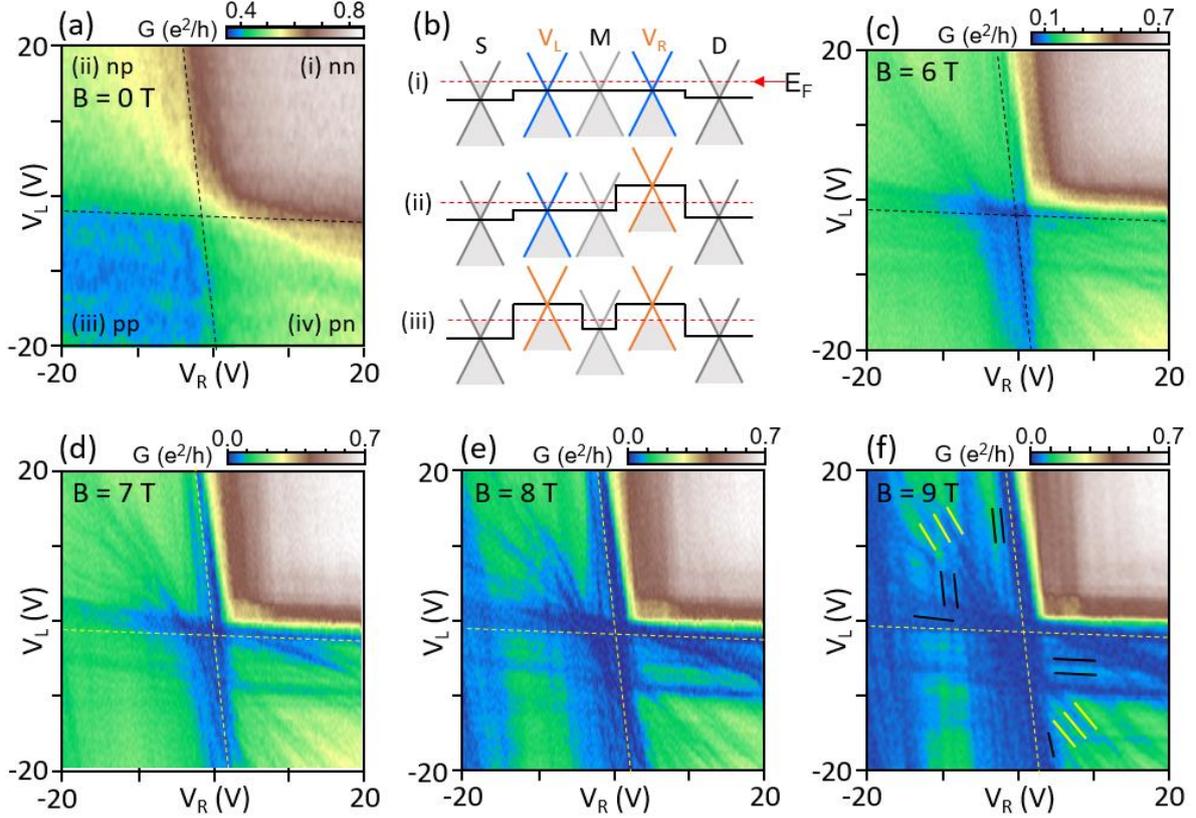

**Figure 2.** (a) Differential conductance map measured as a function of $V_L$ and $V_R$ at $B = 0$ T and $T = 300$ mK. The four regions are labeled according to the type of carrier doping, n– or p–type, of the regions on the left and right, controlled by the respective bottom gates. (b) Schematic illustration of the energy band diagram for unipolar and bipolar regimes. S, D and M represent the source-drain contacts and doping of the supporting part (gray color), respectively, while the segments over the two bottom gates are controlled by $V_L$ and $V_R$ (blue color for n–type and orange color for p–type). The entire nanowire is slightly n–doped. (i) Energy band diagram when both segments over the two bottom gates are tuned to n–type. (ii) Energy band diagram when the gate segment on the right is tuned to p–type, creating an n–p–n cavity. (iii) Both of the bottom gates are tuned to the p–type, creating double n–p–n cavities. Conductance map measured at (c) $B = 6$ T, (d) $B = 7$ T, (e) $B = 8$ T and (f) $B = 9$ T. The conductance in the bipolar regime and p–p regime decreases drastically because of the suppression of the Klein tunneling

as the magnetic field increases. A conductance plateau develops in the n–n regime as the magnetic field increases.

We measure the conductance maps as a function of magnetic field as displayed in Fig. 2c–f. As the magnetic field increases, the boundary of the p–n junction sharpens, showing an abrupt conductance interface along the dashed lines. It is known that the Dirac Fermions can penetrate the p–n junctions with high transmission probability for normal incident angles (known as Klein tunneling),[29-31] resulting in smooth conductance changes (smooth p–n junction) at the p–n junction interface as experienced in the measurement shown in Fig. 2a. However, the situation changes dramatically under high magnetic fields. The Dirac Fermions experience a Lorentz force leading to a cyclotron motion under magnetic fields. Consequently, the trajectories of the electrons at the p–n junction are bent, which reduces the conductance (suppressing the Klein tunneling) in the p–n regime. This results in a more abrupt change of the conductance when entering the p–n regime. At high magnetic fields, the device shows strong conductance suppression and even conductance oscillations in the bipolar (n–p and p–n) and p–p regime. At $B = 9$ T, the conductance oscillations are enhanced significantly in these regimes. This can be attributed to Coulomb oscillations due to quantum dot formation, as shown in Fig. 3c. It should also be noted that clear conductance plateaus appear in the n–n regime at high magnetic fields.

To further investigate the Coulomb oscillations in the p–n (or n–p) regime and the conductance plateaus in the n–n regime at high magnetic fields, we measure the conductance as a function of $V_R$ along direction A (red dashed arrow) in Fig. 3a. Figure 3b shows conductance line cuts along the direction A in Fig. 3a for different magnetic fields, which are plotted with an offset in conductance for clarity. The graph exhibits clear conductance plateaus in the n–n regime ($V_R > \sim 0$) indicated by (i) at higher magnetic fields and the Coulomb oscillations in the p–n regime ($V_R < \sim 0$). We measure the 2D conductance map as a function of $V_{SD}$ and $V_R$ along the line cut A in Fig. 3a and the result is displayed in Fig. 3c. For $V_R < \sim 0$ (in the p–n regime), Coulomb blockade diamonds are observed, showing that the device behaves as a quantum dot. Here, the device forms an n–p–n cavity in the bipolar regime (p–n and n–p) with neighboring n–type leads, which cannot confine carriers in the cavity at $B = 0$ T because of the Klein tunneling. With increasing magnetic field, Klein tunneling is suppressed, thus the cavity starts to behave as a quantum dot. However, the conductance diamonds in Fig. 3c are irregular, indicative of multiple quantum dot behavior. This is further confirmed by the different slopes of the conductance lines marked by the black and yellow lines in Fig. 3a. There are three slopes that suggesting that a quantum dot forms over the left gate, right gate and one in the middle between the two gates. The black lines in the p–n regime are parallel to the Dirac node line (the vertical and horizontal dashed white lines), showing that a quantum dot is mostly controlled by $V_R$ or $V_L$. However, the yellow conductance lines have a different slope along a direction diagonal to $V_L$ and $V_R$, indicating that another quantum dot is formed in-between $V_L$ and $V_R$.

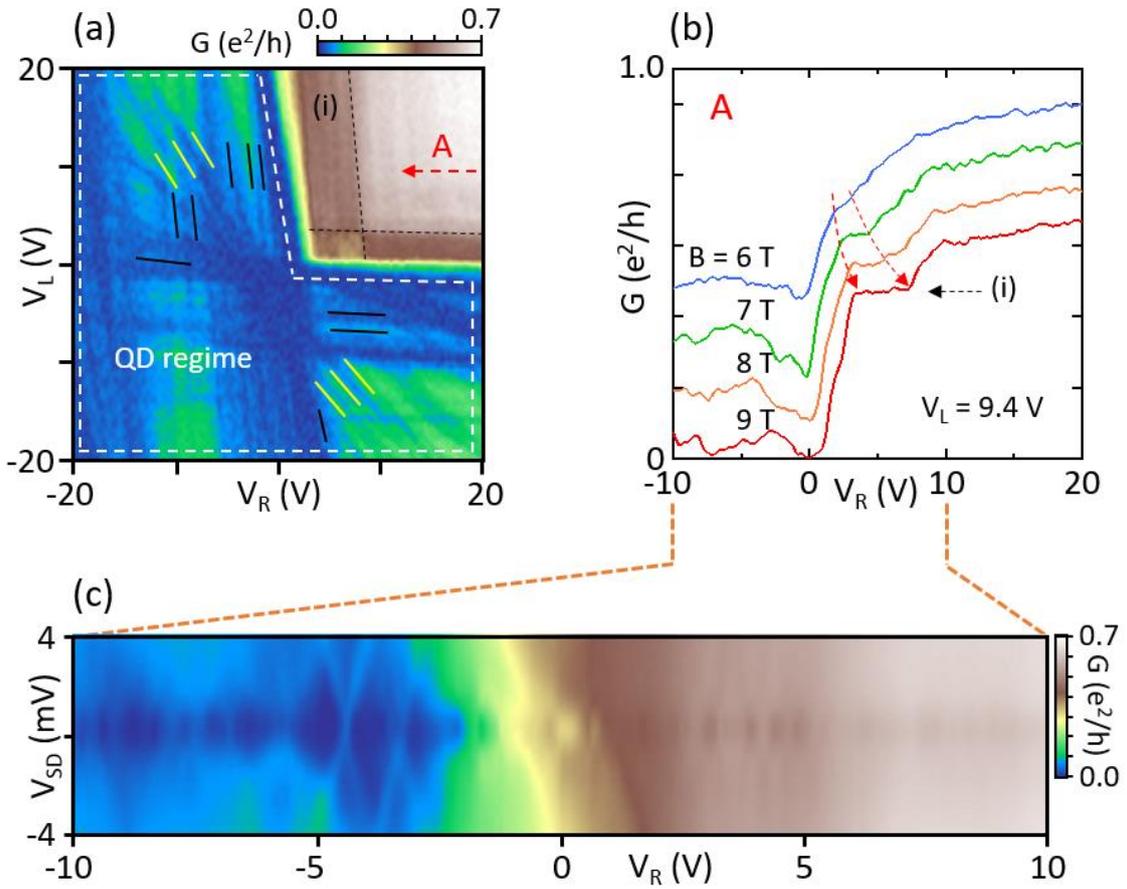

**Figure 3.** Conductance map measured as a function of $V_L$ and $V_R$ at $B$ = 9 T. Coulomb oscillations are observed in the bipolar and p–p regimes, indicating that the device behaves as a quantum dot (enclosed region by white dashed lines). The different slopes of conductance lines marked by black and yellow lines indicate multiple quantum dot behavior. In the n–n regime, the device shows a conductance plateau, indicated by (i). (b) Conductance cuts as a function of $V_R$ taken at $V_L$ = 9.4 V, along the direction A (red dashed arrow) in (a) at different magnetic fields. Conductance cuts are plotted with offset (~0.078 $e^2/h$) between the individual

traces for clarity. (c) 2D conductance plot as a function of $V_{SD}$ and $V_R$ measured along direction A in (a). The Coulomb blockade diamonds are observed for $V_R < -3$ V in the p–n junction. 2D conductance plot as a function of $V_{SD}$ and $V_L$ shows a similar behavior (not shown here).

For $V_R > \sim 0$ (in the n–n regime), the conductance increases abruptly and shows well-defined plateaus in the vicinity of 0.5 $e^2/h$ (point (i) in Fig. 3a and b). A clearly developing plateau is observed as the magnetic field increases from $B = 6$ T to 9 T as indicated by the dashed red arrows in Fig. 3b. This can be attributed to a quantum Hall plateau that originates from the formation of Landau levels in the bulk and quantized states along the edges, so called-edge states. Recently, quantum Hall plateaus have been observed in $Cd_3As_2$ nano-plates[17-19] and thin films.[21-24] In a graphene monolayer, the quantum Hall conductance plateaus are observed at $g$ $(n + ½)$ $e^2/h$, where $g$ is the degeneracy, ½ is the Berry's phase and $n$ is an integer i.e. $n = 0, 1, 2, \ldots$. For monolayer graphene, the value of the degeneracy $g$ is 4, which originates from the spin and valley degrees, thus the filling factors were found at $v = g (n + ½) = 2, 6, 10, \ldots$.[32-35] Despite the degeneracy of $Cd_3As_2$ being the same as that of graphene, the quantum Hall plateaus in $Cd_3As_2$ have been observed at $v \times e^2/h$, where the filling factor $v$ has been found to be different sets of integer values. In particular, even filling factors $v = 2, 4, 6$, were found due to the quantum confinement induced bulk subbands in the thin film.[22-24] In contrast, the only odd-integer filling factors $v = 1, 3, 5$, were observed due to the topological surface states.[19] Recently Zhang et al. observed continuous integer values (both odd and even) $v = 1, 2, 3, \ldots$ based on the Weyl orbits in $Cd_3As_2$ nano-plates.[18] However, in our Device A, the plateaus are not quantized at an integer multiple of $e^2/h$, as observed in the other $Cd_3As_2$ devices mentioned above. In our work, the observed conductance value is smaller than 1 $e^2/h$ at the plateau and this is mostly due to the contact resistance (The typical contact resistance ($R_C$) for nanowires is $\sim 3–10$ k$\Omega$, also see Supporting Information, Fig. S4.).

We further perform the same experiment on a device with a $Cd_3As_2$ nano-ribbon (Device B) of which the thickness and width of the nano-ribbon are $\sim 60$ and $\sim 300$ nm, respectively. The differential conductance is measured as a function of $V_L$ at $B = 8$ T and $T = 3.5$ K as shown in Fig. 4a. The device shows a clear conductance plateau near $1e^2/h$. This can be understood, since the contact resistance to a nano-ribbon is much lower than that of the nanowire. The second plateau is not clearly observed because the magnetic field is not sufficiently strong. Figure 4b shows the evolution of conductance plateau as the magnetic field is increased. Although the exact reason of the 1 $e^2/h$ plateau in our work remains unclear yet, a possible scenario involves the quantum Hall effect based on the Weyl orbits.[18]

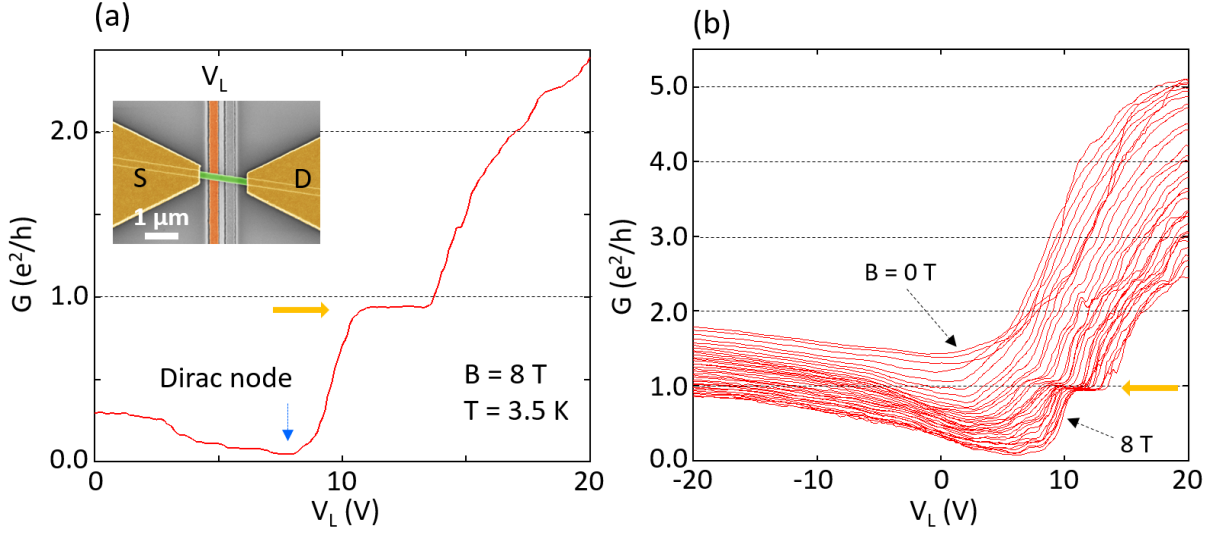

**Figure 4.** (a) Conductance measured in a nano-ribbon (Device B) as a function of $V_L$ at $B = 8$ T and $T = 3.5$ K. A clear conductance plateau is observed at approximately $1e^2/h$, as indicated by the orange arrow. Inset: SEM image of the device. The thickness and width of the nano-ribbon are 60 and 300 nm, respectively. In this device, the right gate is not operational. (b) Conductance measured as a function of the magnetic field. As the magnetic field increases, a clear conductance plateau develops at approximately $1\ e^2/h$ (the orange arrow).

In summary, we have fabricated suspended 3D DSM $Cd_3As_2$ nanowire devices with two recessed bottom gates and realize DSM p–n junctions. We have observed four different conductance regimes, n–n, p–p, n–p and p–n, depending on the two bottom gate voltages. In the bipolar regimes, the device operates as a quantum dot confined within two p–n junctions at high magnetic fields by suppressing the Klein tunneling. In the homogeneous doping regime (n–n), the device shows a clear conductance plateau owing to the quantum edge channel at high magnetic fields. Our results demonstrate that DSM nanowires and nano-ribbons could be used for future quantum devices such as single and double quantum dots. Moreover, the electron optics that was demonstrated in graphene p–n junctions[36] could also be realized in 3D DSM.

# Conflicts of interest

There are no conflicts to declare.

# Acknowledgements

MJ thanks H.-S. Sim, M.-S. Choi and H. C. Park for helpful discussions. This work is supported by the Mid-career Researcher Program (NRF-2017R1A2B4007862) and DGIST R&D Program of the Ministry of Science, ICT, Future Planning (19-NT-01). Research at University of Tokyo was supported by Grant-in-Aid from JSPS (No. 17H01038), MEXT Grant-in-Aid for Scientific Research on Innovative Areas "Science of hybrid quantum systems" (No.15H05868). M.B.A.J. would like to acknowledge the MOE Tier I (NUS Grant No. R-263-000-B98-112, NUS Grant No. R-263-000-D66-114), MOE Tier II MOE2013-T2-2-125 (NUS Grant No. R-263-000-B10-112) grants, MOE Tier II MOE2018-T2-2-117 (NUS Grant No. R-398-000-092-112) grants and NRF-CRP12-2013-01 (NUS Grant No. R-263-000-B30-281) for financial support. N.N. is supported by JST CREST Grant No. JPMJCR1874 and JPMJCR16F1, Japan, and JSPS KAKENHI Grant No. 18H03676 and 26103006. X.-X.Z. was partially supported by NSERC and CIfAR. CS acknowledges financial support from the ERC projects QUEST & TopSupra and the Swiss National Science Foundation (SNF) through various grants, including the NCCR-QSIT and the Swiss Nanoscience Institute. MJ thanks Hyeon Jeong Lee, Hwan Soo Jang and Bong Ho Lee at CCRF of DGIST for device fabrication and helpful discussions.

# Supporting Information

p-n junctions in Device C and D, conductance vs gate voltage for Device A and C, contact resistance extracted from 2- and 4-terminal measurements

# Supporting Information

# Controllable p–n junctions in three-dimensional Dirac semimetal Cd$_3$As$_2$ nanowires

Janice Ruth Bayogan,[†,‡,¶] Kidong Park,[§,¶] Zhuo Bin Siu,[∥,¶] Sung Jin An,[†] Chiu-Chun Tang,[⊥] Xiao-Xiao Zhang,[#,▽] Man Suk Song,[†,‡] Jeunghee Park,[§] Mansoor B. A. Jalil,[∥] Naoto Nagaosa,[#,@] Kazuhiko Hirakawa,[*,⊥] Christian Schönenberger,[*,△] Jungpil Seo[*,†,‡] and Minkyung Jung[*,‡]

† *Department of Emerging Materials Science, DGIST, Daegu 42988, Korea*

‡ *DGIST Research Institute, DGIST, Daegu 42988, Korea*

§ *Department of Chemistry, Korea University, Sejong 339-700, Korea*

∥ *Electrical and Computer Engineering, National University of Singapore, Singapore 117576, Republic of Singapore*

⊥ *Institute for Nano Quantum Information Electronics, IIS, The University of Tokyo, Tokyo 153-8505, Japan*

# *Department of Applied Physics, The University of Tokyo, Tokyo 113-8656, Japan*

▽ *Department of Physics and Astronomy & Stewart Blusson Quantum Matter Institute, University of British Columbia, Vancouver, BC, V6T 1Z4, Canada*

@ *RIKEN Center for Emergent Matter Science (CEMS), Saitama 351-0198, Japan*

△ *Department of Physics, University of Basel, Klingelbergstrasse 82, CH-4056 Basel, Switzerland*

¶ *J.R.B., K.P. and Z.B.S contributed equally to this work.*

E-mail: minkyung.jung@dgist.ac.kr; jseo@dgist.ac.kr; Christian.Schoenenberger@unibas.ch; hirakawa@iis.u-tokyo.ac.jp

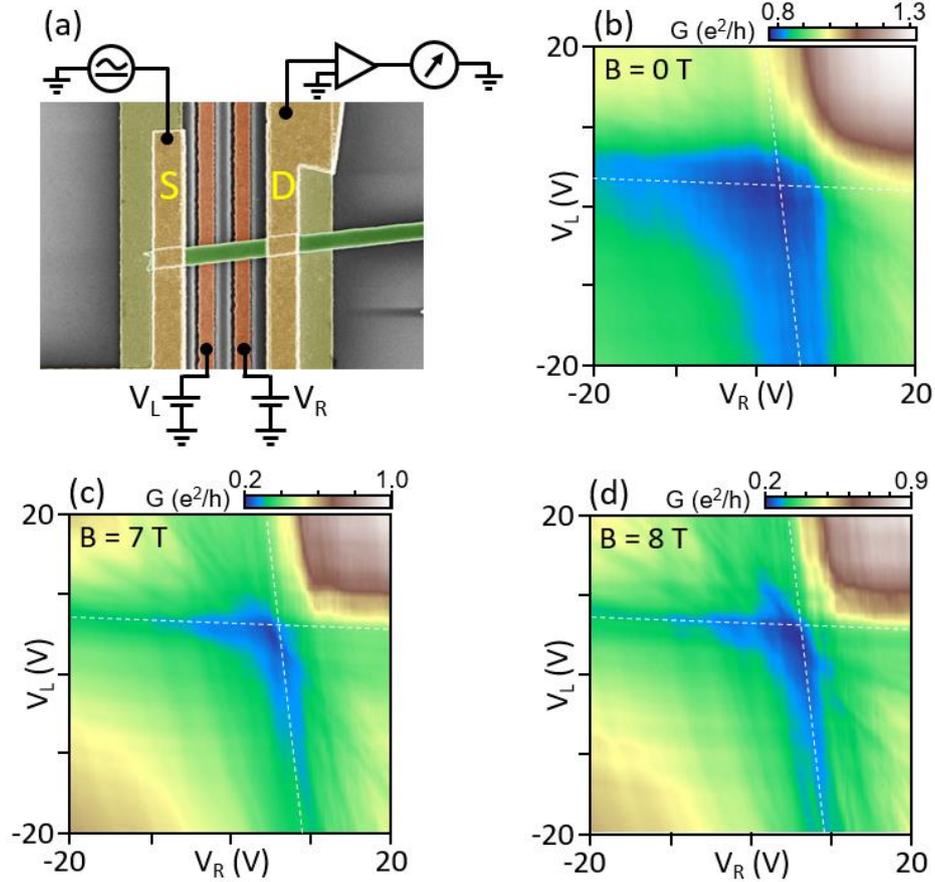

**Figure S1.** (a) SEM image of fully suspended $Cd_3As_2$ nanoribbon p-n junction device (device C). All the fabrication procedure is same with device A, except for the supporting bar (SB) as indicated by the yellow arrow. The nanoribbon is transferred onto the supporting bars and contacted with 10/100 nm of Ti/Au, so that the entire nanoribbon is fully suspended. The channel width and thickness of the nanoribbon are ~ 300 nm and ~ 80 nm, respectively. The device is measured at 2 K. Conductance is measured as a function of $V_L$ and $V_R$ for (b) $B = 0$ T, (c) $B = 7$ T and (d) $B = 8$ T.

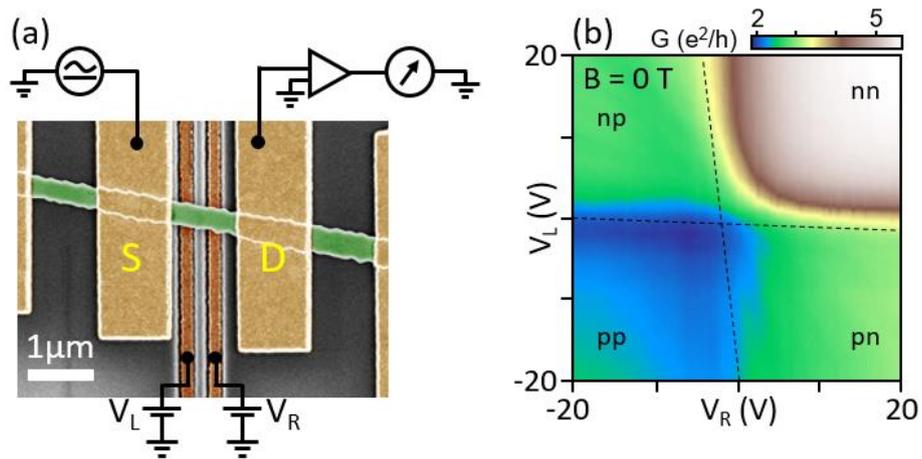

**Figure S2.** (a) SEM image and measurement setup of $Cd_3As_2$ nanoribbon p-n junction device (device D). The channel width and thickness of the nanoribbon are ~ 300 nm and ~ 80 nm, respectively. (b) Conductance is measured as a function of $V_L$ and $V_R$ for $B = 0$ T and $T = 2$ K.

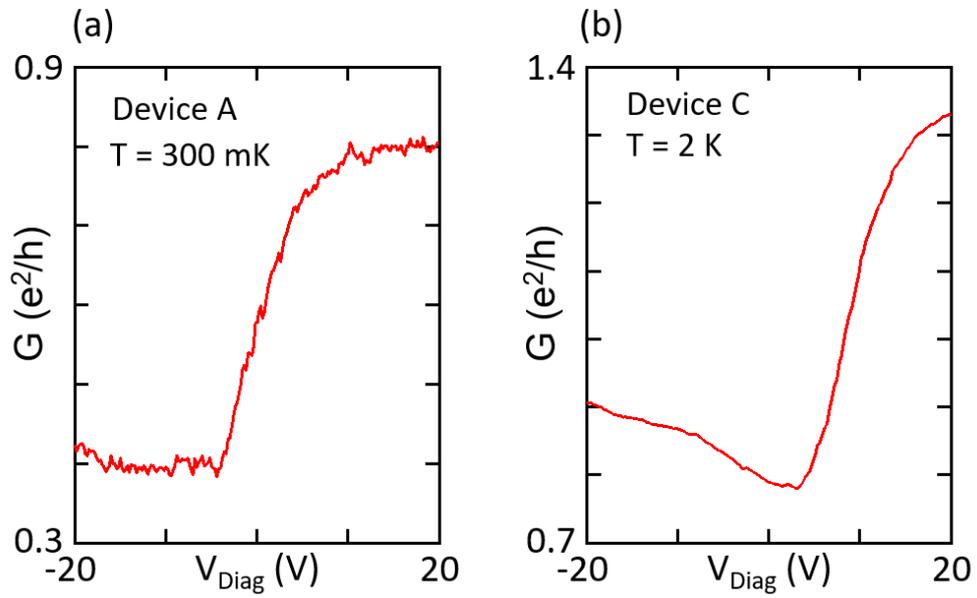

**Figure S3**. Conductance as a function of gate voltage ($V_L = V_R$) measured in (a) Device A at $T = 300$ mK and (b) Device C at $T = 2$ K for $B = 0$ T.

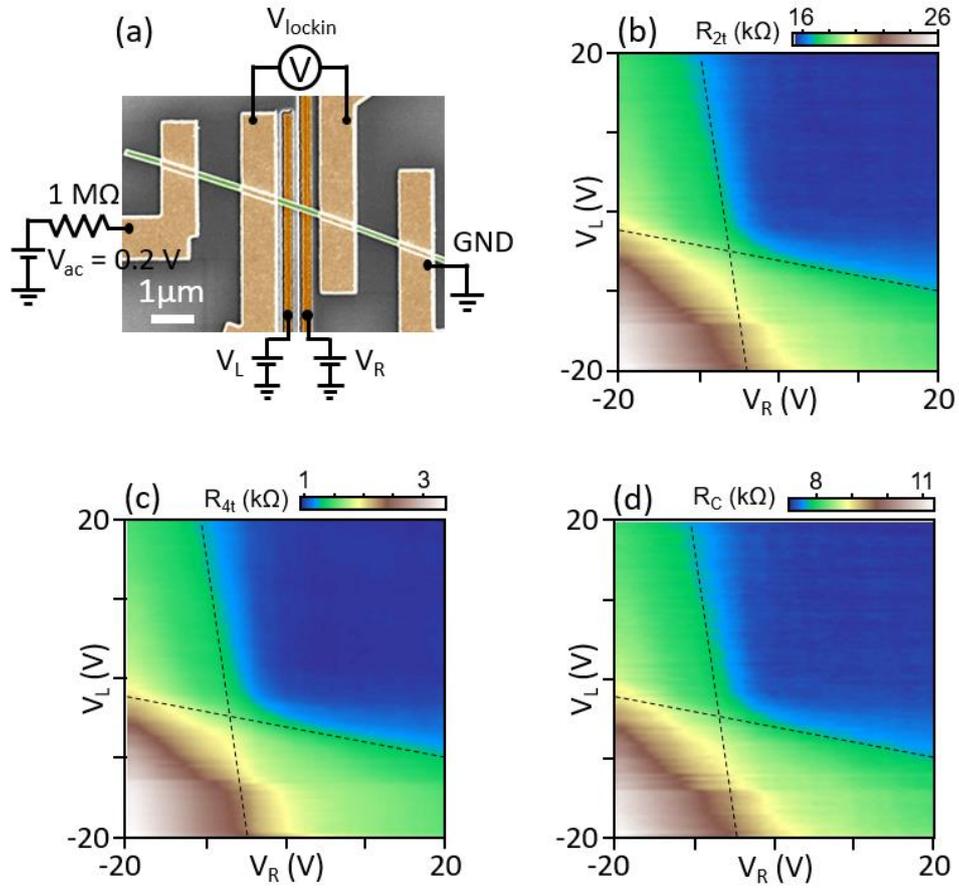

**Figure S4**. (a) SEM image of Device E with 4-terminal measurement setup. Resistance as a function of $V_L$ and $V_R$ measured with (b) 2-terminal measurement and (c) 4-terminal measurement setup. (d) Contact resistance is extracted from the 2- and 4-terminal measurements. The contact resistance is defined as $R_C = (R_{2t} - R_{4t})/2$.